\documentclass[english,aps,pra,twocolumn,showpacs,floatfix,superscriptaddress]{revtex4-1}
\usepackage{amsmath}
\usepackage{graphicx}
\usepackage{amssymb}
\usepackage{color}

\makeatletter

\usepackage{babel}
\makeatother

\begin{document}
\newcommand{\bra}[1]{\left\langle #1\right|}
\newcommand{\ket}[1]{\left|#1\right\rangle }
\newcommand{\braket}[2]{\left\left \langle#1|#2\right\rangle }
\newcommand{\tr}{\mathrm{tr}}
\newcommand{\commute}[2]{\left[#1,#2\right]}
\newcommand{\anticommute}[2]{\left\{#1,#2\right\}}
\newcommand{\expect}[1]{\left\langle #1\right\rangle }
\newcommand{\sans}[1]{\mathsf{#1}}
\newcommand{\jcomment}[1]{\emph{\color[rgb]{0,0,0.6}(#1)}}
\newcommand{\bcomment}[1]{\emph{\color[rgb]{0.6,0,0}(#1)}}


%
%

\title{Hybridization and spin decoherence in heavy-hole quantum dots}

\date{\today}

\author{Jan Fischer}
\affiliation{Department of Physics, University of Basel, Klingelbergstrasse 82,
CH-4056 Basel, Switzerland}
\affiliation{Institut f{\"u}r Theoretische Physik, Universit{\"a}t Regensburg, D-93040 Regensburg, Germany}
\author{Daniel Loss}
\affiliation{Department of Physics, University of Basel, Klingelbergstrasse 82,
CH-4056 Basel, Switzerland}

\begin{abstract}
We theoretically investigate the spin dynamics of a heavy hole confined to an unstrained III-V semiconductor
quantum dot and interacting with a narrowed nuclear-spin bath.
We show that band hybridization leads to an exponential decay of hole-spin superpositions
due to hyperfine-mediated nuclear pair flips, and that the accordant
single-hole-spin decoherence time $T_2$ can be tuned over many orders of magnitude by changing external 
parameters. In particular, we show that, under experimentally accessible conditions, it is possible
to suppress hyperfine-mediated nuclear-pair-flip processes so strongly that hole-spin quantum dots 
may be operated beyond the `ultimate limitation' set by the hyperfine interaction which is present in
other spin-qubit candidate systems.
\end{abstract}


\maketitle

Heavy holes (HH) confined to semiconductor quantum dots (QD) have attracted rapidly growing attention
over the last years for their potential applicability in spintronics and as qubits for
quantum information processing devices. The spin states of confined HHs feature very long spin relaxation 
times \cite{bulaev:2005, heiss:2007, trif:2009} and are suspected to be robust against spin decoherence.
Typically, for spin qubits operated at sub-Kelvin temperatures, the main source of decoherence is the interaction with the
nuclear spins residing in the host material and the inhomogeneous broadening of the nuclear magnetic field
(Overhauser field) \cite{petta:2005}. 
For HHs, the form of the nuclear-spin interaction is predominantly Ising-like
\cite{fischer:2008}, in contrast to the Heisenberg-type interaction of electrons.
Hole-spin QDs in $p$-type GaAs/AlGaAs heterostructures have already been realized experimentally
and operated in the few-hole regime \cite{komijani:2008, csontos:2010, klochan:2010}.
Experiments in self-assembled InGaAs quantum dots have shown the possibility to
initialize and read out the spin state of a HH with high fidelity \cite{gerardot:2008},
and ensemble-spin decoherence times $T_2^*$ on the order of hundreds of nanoseconds have been measured
\cite{brunner:2009}.

Several possibilities to suppress decoherence due to inhomogeneous broadening have been proposed,
one of which is to prepare the nuclear spins in a so-called narrowed or frequency-focused state
\cite{coish:2004, klauser:2006, stepanenko:2006}, where the bath is prepared in an eigenstate
of the Overhauser operator (see text below Eq. \eqref{eq:GME}).
On the experimental side, enormous progress has been achieved in preparing such narrowed states 
\cite{greilich:2006, greilich:2007, reilly:2008, greilich:2009, latta:2009, xu:2009, vink:2009, bluhm:2010},
which have been shown to persist over astonishingly long timescales exceeding hours \cite{greilich:2007}.
For electrons interacting with a narrowed nuclear bath, spin decoherence happens due to 
nuclear pair flip processes induced by the transverse hyperfine interaction, and the associated single-spin 
decoherence time $T_2$  can be several orders of magnitude 
longer than the ensemble-spin decoherence time $T_2^*$ \cite{coish:2008, cywinski:2009}.
For HHs, with their predominantly Ising-like coupling to nuclear spins \cite{fischer:2008}, this transverse interaction
(perpendicular to the Ising axis) can be expected to be very small, potentially leading to very
long single-hole-spin decoherence times $T_2$. 

In this Letter, we study the spin dynamics of a HH confined to a III-V semiconductor QD
and interacting with a narrowed nuclear-spin bath. 
We show that band hybridization leads to non-Ising (transverse) terms in the hyperfine Hamiltonian, 
whose magnitude depends on the geometry of the QD. This transverse coupling induces nuclear pair-flip processes, 
leading to fluctuations of the Overhauser field and to exponential single-hole-spin
decoherence. We show that for typical unstrained quantum dots the associated timescale
$T_2$ has a lower bound on the order of tens of microseconds and that it can be tuned over many orders of 
magnitude by changing external parameters such as the applied magnetic field. 
Thus, it is in principle possible to operate hole-spin QDs in a regime where the hyperfine interaction is
practically switched off and where other decoherence mechanisms, such as nuclear dipole or spin-orbit interactions, 
will become relevant and, hence, experimentally observable.

We start from the $8 \times 8$ Kane Hamiltonian describing states in the conduction band (CB), heavy-hole (HH),
light-hole (LH) and split-off (SO) bands of bulk III-V semiconductors
(see Appendix C of Ref. \cite{winkler}). The Kane Hamiltonian can be `folded down' to an effective
$2 \times 2$ Hamiltonian whose eigenstates describe the spin states in the band of interest, 
where the admixture of neighboring bands is taken into account perturbatively \cite{winkler}. 
Using this procedure, we find the following hybridized HH pseudospin states (see Appendix \ref{appendix1}):
\begin{multline}
  \label{eq:spinstates}
  |\Psi_\pm \rangle \simeq \mathcal{N} \Bigl( |u_{\mathrm{HH}}^{\pm}; \phi_{\mathrm{HH}}^{00} \rangle |\pm_{\mathrm{HH}} \rangle\\
  \mp \lambda_{\mathrm{CB}} \, |u_{\mathrm{CB}}^{\pm}; \phi_{\mathrm{CB}}^{0\pm} \rangle |\pm_{\mathrm{CB}} \rangle
  \pm \lambda_{\mathrm{LH}} \, |u_{\mathrm{LH}}^{\pm}; \phi_{\mathrm{LH}}^{1\pm} \rangle |\pm_{\mathrm{LH}} 
  \rangle \Bigr).
\end{multline}
Here, we have assumed a parabolic confinement potential 
defining a QD with lateral and perpendicular confinement lengths $L$ and $a_z$, respectively, 
and $\mathcal{N}$ enforces proper normalization of the wavefunctions.
The condition for the validity of Eq. \eqref{eq:spinstates}
is given by $a_z \ll L$, which is needed for the perturbation expansion on the Kane Hamiltonian.
The amount of CB and LH admixture is determined by $\lambda_{\mathrm{CB}} = i \beta_{\mathrm{CB}} P/\sqrt{2} L E_g$
and $\lambda_{\mathrm{LH}} = \sqrt{3}  \beta_{\mathrm{LH}} \gamma_3 a_z L / 2 \sqrt{2} \gamma_2 (L^2-a_z^2)$, respectively,
where $P$ is the interband momentum, $E_g$ is the band gap, $\gamma_{2,3}$ are Luttinger parameters,
and $\beta_{\mathrm{CB}}$, $ \beta_{\mathrm{LH}}$
account for the difference in effective masses between the bands (see Appendix \ref{appendix1}).

Near the $\Gamma$-point, the spin-orbit-coupled states can be approximated by 
$|u_{\mathrm{CB}}^{\pm}\rangle |\pm_{\mathrm{CB}}\rangle \simeq {|s\rangle |\uparrow, \downarrow\rangle}$,
$|u_{\mathrm{HH}}^{\pm}\rangle |\pm_{\mathrm{HH}}\rangle \simeq {|p_\pm\rangle |\uparrow, \downarrow \rangle}$,
$|u_{\mathrm{LH}}^{\pm}\rangle |\pm_{\mathrm{LH}}\rangle \simeq ({\sqrt{2} |p_z\rangle |\uparrow, \downarrow \rangle} 
\mp {|p_\pm\rangle |\downarrow, \uparrow \rangle})/\sqrt{3}$,
in terms of $s$- and $p$-symmetric Bloch states ($|p_\pm\rangle = |p_x\rangle \pm i |p_y \rangle$)
and real-spin states $|\uparrow, \downarrow \rangle$ 
with respect to the growth axis \cite{winkler}.
The envelope functions appearing in Eq. \eqref{eq:spinstates} are defined via their position representations
$\langle \mathbf{r} | \phi_\alpha^{ij} \rangle = \phi_\alpha^{i \perp}(z) \phi_\alpha^{j \|}(x,y)$ ($i=0,1$, $j=0,\pm$), 
where $\phi_\alpha^{0 \|}(x,y)=\phi_\alpha^0(x) \phi_\alpha^0(y)$, $\phi_\alpha^{\pm \|}(x,y) = 
(\phi_\alpha^1(x) \phi_\alpha^0(y) \pm i \phi_\alpha^0(x) \phi_\alpha^1(y))/\sqrt{2}$,
and $\phi_\alpha^n(x)$ is the $n^\mathrm{th}$ harmonic-oscillator eigenfunction in band $\alpha$.
Due to terms appearing in the Kane Hamiltonian which are linear in the crystal momentum $\mathbf{k}$ and which
couple neighboring bands, the admixture of CB and LH states features \emph{excited-state} envelope functions.
This has profound physical consequences which will be discussed below.
The split-off-band contribution to the HH states is very small and has thus been neglected in Eq. \eqref{eq:spinstates}.

There are three interactions that couple an electron (or HH) to the spins of the surrounding nuclei:
the Fermi contact interaction $h_1^k$, the anisotropic hyperfine interaction $h_2^k$, and the
coupling of orbital angular momentum to the nuclear spins $h_3^k$, which read
(setting $\hbar=1$) \cite{stoneham}:
\begin{align}
  \label{ham:contact}
  h_1^k &= \frac{\mu_0}{4 \pi} \> \frac{8 \pi}{3} \> \gamma_S \gamma_{j_k} \>
  \delta(\mathbf{r}_{k}) \> \tilde{\mathbf{S}} \cdot \mathbf{I}_k,\\
  \label{ham:anisotropic}
  h_2^k &= \frac{\mu_0}{4 \pi} \> \gamma_S \gamma_{j_k} \> \frac{3 (\mathbf{n}_k \cdot 
  \tilde{\mathbf{S}}) (\mathbf{n}_k \cdot \mathbf{I}_k) - \tilde{\mathbf{S}} \cdot \mathbf{I}_k}
  {r_{k}^3 (1+d/r_{k})},\\
  \label{ham:angular}
  h_3^k &= \frac{\mu_0}{4 \pi} \> \gamma_S \gamma_{j_k} \> \frac{\mathbf{L}_k \cdot 
  \mathbf{I}_k} {r_{k}^3 (1+d/r_{k})}.
\end{align}
Here, $\gamma_S=2\mu_B$, $\gamma_{j_k}=g_{j_k} \mu_N$, $\mu_B$ is the Bohr magneton,
$g_{j_k}$ is the nuclear g-factor of isotopic species $j_k$ at lattice site $k$,
$\mu_N$ is the nuclear magneton, $\mathbf{r}_{k} = \mathbf{r} - \mathbf{R}_k$ 
is the electron-spin position operator relative to the $k^\mathrm{th}$ nucleus with spin $\mathbf{I}_k$, 
$d \simeq Z \times 1.5 \times 10^{-15} \, \mathrm{m}$, 
$Z$ is the charge of the nucleus, and $\mathbf{n}_k = \mathbf{r}_{k}/ r_{k}$. 
$\tilde{\mathbf{S}}$ and $\mathbf{L}_k = \mathbf{r}_k \times \mathbf{p}$ denote the spin ($m_{\tilde{S}} = \pm 1/2$) and 
orbital angular-momentum operators of the electron, respectively.

In order to derive an effective spin Hamiltonian for the HH,
we take matrix elements $\langle \Psi_\tau | h_1^k + h_2^k + h_3^k | \Psi_{\tau'} \rangle = H_{\tau \tau'}$ ($\tau, \tau' = \pm$)
with respect to the hybridized HH wavefunctions \eqref{eq:spinstates}. Due to the $\delta$-function
in Eq. \eqref{ham:contact}, only the CB admixture contributes to the Fermi contact interaction,
since $p$-states vanish at the positions $\mathbf{R}_k$ of the nuclei. On the other hand, the terms in
Eq. \eqref{eq:spinstates} associated with HH and LH states
contribute to matrix elements of Eqs. \eqref{ham:anisotropic} and \eqref{ham:angular}, while the
CB admixture does not contribute due to symmetry ($h_2^k$) and vanishing orbital angular momentum ($h_3^k$).
Adding up all contributions, and taking into account a Zeeman term due to a magnetic field $B$ along the
$z$-direction, we find the following effective spin Hamiltonian describing the
hole-nuclear-spin interactions:
\begin{equation}
  \label{eq:holehamiltonian}
  H = (b+h^z) S^z + \frac{1}{2} (h^+ S^- + h^- S^+).
\end{equation}
Here, $b=g_h \mu_B B$ is the Zeeman energy of the HH, $g_h \simeq 2$ is the HH g-factor along the magnetic-field
direction $z$, $\mu_B$ is the Bohr magneton, and $\mathbf{S}$ is the HH pseudospin-$1/2$ operator.
The Overhauser-field components are defined by
$h^z = \sum_k A_k^z I_k^z$ and $h^\pm = \sum_k A_k^\pm I_k^\pm$ ($I_k^\pm = I_k^x \pm i I_k^y$),
where $A_k^z$ and $A_k^\pm$ denote the longitudinal and transverse hyperfine coupling of the HH to
the $k^\mathrm{th}$ nuclear spin, respectively.
The flip-flop terms in Eq. \eqref{eq:holehamiltonian} couple the HH pseudospin ($\pm 3/2$) states
through admixture with CB and LH pseudospin ($\pm 1/2$) states, such that flip-flop processes with $I=1/2$ 
nuclear spins preserve the total angular momentum.

The hybridized states in Eq. \eqref{eq:spinstates} are predominantly HH-like. 
In Ref. \cite{fischer:2008} it has been shown that taking matrix elements of the Hamiltonians
\eqref{ham:contact}-\eqref{ham:angular} with respect to pure HH states (i.e., neglecting
band hybridization) results in an Ising Hamiltonian $h^z S^z$. 
The longitudinal coupling constants are thus dominated by the HH contribution, $A_k^z \simeq A_{k,\mathrm{HH}}^z$,
and the transverse (non-Ising) terms in Eq. \eqref{eq:holehamiltonian} are only due to hybridization with CB and LH states,
$A_k^\pm = A_{k,\mathrm{CB}}^\pm + A_{k,\mathrm{LH}}^\pm$.
Explicitly, the longitudinal and transverse coupling constants are given by
$A_{k,\mathrm{HH}}^z \simeq A_{\mathrm{HH}}^{j_k} v_0 |\phi_0(z_k)|^2 |\phi_0 (x_k,y_k)|^2$,
$A_{k,\mathrm{CB}}^\pm \simeq A_{\mathrm{CB}}^{j_k} v_0 |\phi_0(z_k)|^2 \phi^*_\pm (x_k,y_k) \phi_\mp(x_k,y_k)$, and
$A_{k,\mathrm{LH}}^\pm \simeq A_{\mathrm{LH}}^{j_k} v_0 |\phi_1(z_k)|^2 \phi^*_\mp (x_k,y_k) \phi_\pm(x_k,y_k)$, respectively, 
where $v_0$ is the volume occupied by one nucleus and $A_{\alpha}^{j_k}$ is the hyperfine coupling strength of
isotope $j_k$ associated with band $\alpha$.
Introducing the average $A_\alpha = \sum_j \nu_j A_\alpha^j$, where $\nu_j$ denotes the abundance of isotope $j$, we estimate 
$A_{\mathrm{HH}} \simeq -13 \mu e \mathrm{V}$ \cite{fischer:2008}, $A_{\mathrm{CB}} \simeq 0.15 \mu e \mathrm{V}$, and 
$A_{\mathrm{LH}} \simeq 0.05 \mu e \mathrm{V}$ for a GaAs QD with $L=10 \mathrm{nm}$ and $a_z=2 \mathrm{nm}$.
In contrast to the interaction of an electron with nuclear spins,
the hole-nuclear-spin interaction given in Eq. \eqref{eq:holehamiltonian} is highly anisotropic.

We now study the dynamics of the transverse spin component $S^+$ describing the coherence of the HH pseudospin states.
To this end, we use the Nakajima-Zwanzig master equation \cite{coish:2004}
\begin{equation}
  \label{eq:GME}
  \langle \dot{S}^+ \rangle_t = i \omega_n \langle S^+ \rangle_t -i \int_0^t dt' \,
  \Sigma(t-t') \langle S^+ \rangle_{t'},
\end{equation}
where $\omega = b+h^z$, $\omega |n\rangle = \omega_n |n\rangle$, and $|n\rangle$ denotes a narrowed state of the 
nuclear-spin system (note that for a non-narrowed bath, the HH decoherence would be dominated by the
Ising part of Eq. \eqref{eq:holehamiltonian}, as shown in Ref. \cite{fischer:2008}).
$\Sigma(t) = \mathrm{tr} \{ S^+ \hat{\Sigma}(t) S^- |n \rangle \langle n| \}$ is the self-energy (or memory kernel) describing the
transverse-spin dynamics, where $\hat{\Sigma}(t) = -i\mathsf{P}L\mathsf{Q} e^{-iL\mathsf{Q}t} \mathsf{Q}L\mathsf{P}$, 
$\mathsf{P}$ is a projector onto a product state of HH and nuclear spins, $\mathsf{Q}=1-\mathsf{P}$, and 
$L \mathcal{O} = [H, \mathcal{O}]$ for some operator $\mathcal{O}$ acting on the total Hilbert space of HH and nuclear spins 
\cite{coish:2004}.
It is convenient to perform a Laplace transform 
on Eq. \eqref{eq:GME}, yielding an algebraic equation of the form
\begin{equation}
  \label{eq:GME-Laplace}
  S^+(s+i \omega_n) = \frac{\langle S^+ \rangle_0}{s +i \Sigma(s+i \omega_n)}
\end{equation}
in the frame rotating with frequency $\omega_n$.
Eqs. \eqref{eq:GME} and \eqref{eq:GME-Laplace} are exact equations describing, in general, non-Markovian dynamics 
of the transverse HH-spin component. The structure of the self-energy $\Sigma(s)$ is, however, very complex, so we have to
resort to an approximation scheme. The energy scales associated with the transverse coupling
$V=(h^+ S^- + h^- S^+)/2$ are much smaller than those associated with the longitudinal coupling
$H_0 = (b+h^z)S^z$ (see above), and we expand the self-energy in powers of hole-nuclear-spin flip-flop processes
induced by $V$: $\Sigma(s) = \Sigma^{(2)}(s) + \Sigma^{(4)}(s) + \mathcal{O}(V^6)$. 
Odd orders in $V$ vanish because of the Zeeman mismatch between HH and nuclear spins which energetically forbids such processes.
For a nuclear spin $I$ of order
unity, the smallness parameter which controls this expansion is given approximately by $A_\perp/\omega_n$ 
(see Appendix A of Ref. \cite{coish:2004}), where $A_\perp = \sqrt{A_{\mathrm{CB}}^2 + A_{\mathrm{LH}}^2}$.

\begin{figure}[t]
  \centering
  \includegraphics[width=.98\columnwidth]{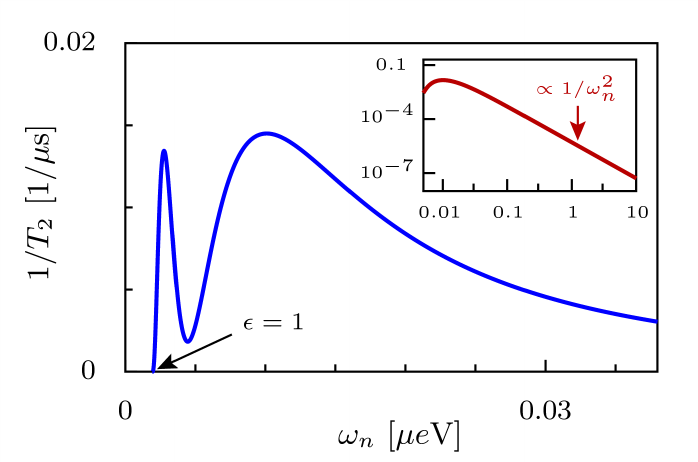}
  \caption{Decoherence rate $1/T_2$ from Eq. \eqref{eq:Gamma} as a function of the HH Zeeman energy
    $\omega_n = g_h \mu_B B + p I A_{\mathrm{HH}}$.
    For $L=10\mathrm{nm}$ and $a_z=4\mathrm{nm}$, we estimate  $N \simeq 7.3 \times 10^4$. 
    Inset: $1/T_2$ for fields up to $1\mathrm{T}$  (axes in the same units as in the main figure).}
  \label{fig:Gamma-of-B}
\end{figure}

We evaluate the second- and fourth-order self-energy contributions explicitly, following the procedure 
described in Ref. \cite{coish:2010}. We find, for a homonuclear system in the frame rotating with frequency
$\omega_n$,
\begin{align}
  \label{eq:sigma2}
  \Sigma^{(2)}(s+i \omega_n) &\simeq - \frac{c_+ + c_-}{4 \omega_n} \sum_k |A_k^\pm|^2,\\
  \Sigma^{(4)}(s+i \omega_n) &\simeq -i \frac{c_+ c_-}{4 \omega_n^2} \sum_{k_1,k_2}
  \frac{|A_{k_1}^\pm|^2 |A_{k_2}^\pm|^2}{s+i (A_{k_1}^z - A_{k_2}^z)}, \label{eq:sigma4}
\end{align}
where the sums run over all nuclear sites $k$. 
We have introduced $c_\pm = I(I+1)-\langle\langle m(m\pm1) \rangle\rangle$, where $I$ is the nuclear spin,
$m=-I, \ldots, I$, and the double angle bracket indicates averaging over the $I_k^z$ eigenvalues $m$ \cite{coish:2004}.

We emphasize that the structure of the
self-energies $\Sigma^{(2)}$ and $\Sigma^{(4)}$ bears some similarity with previous results on
electron-spin decoherence \cite{coish:2010}. However, there are two important differences compared to the
electron case: (i) The appearance of different coupling constants $A_k^z$ and $A_k^\pm$ in Eqs. 
\eqref{eq:sigma2} and \eqref{eq:sigma4} is due to the anisotropy of the hyperfine Hamiltonian 
\eqref{eq:holehamiltonian} and provides an additional smallness factor $A_\perp / A_z \ll 1$
($A_z = |A_{\mathrm{HH}}|$) to the self-energy \eqref{eq:sigma4cont}.
(ii) The spatial dependence of the transverse coupling constants differs from
the longitudinal ones due to the appearance of excited-state envelope functions.
In particular, this means that nuclear spins at the edge of the QD (rather than in its center
as in the electron case) couple most strongly to the HH along the transverse direction -- an effect which 
manifests itself directly in the appearance of a distinct minimum in the decoherence rate $1/T_2$ (see Fig. \ref{fig:Gamma-of-B}).

We now evaluate the second- and fourth-order self-energy in the continuum limit (changing sums to integrals,
see Appendix \ref{appendix2}), following Ref. \cite{coish:2010}.
Since $a_z \ll L$ (see above), we can perform a two-dimensional limit by averaging over the $z$-dependence 
in the hyperfine coupling constants $A_k^z$ and $A_k^\pm$.
From Eq. \eqref{eq:sigma2}, we see that the second-order self-energy $\Sigma^{(2)}$ is purely real, 
leading to no decay but a frequency shift $\Delta \omega = - \mathrm{Re} \Sigma^{(2)}(s+i \omega_n)$, or
\begin{equation}
  \label{eq:deltaomega}
  \Delta \omega = \frac{c_+ + c_-}{16 N} \, \frac{A_\perp^2}{\omega_n},
\end{equation}
where $N$ is the number of nuclear spins enclosed by the envelope function.
The fourth-order self-energy becomes
\begin{align}
  \nonumber
  \Sigma^{(4)}(s+i &\omega_n) \simeq -i \frac{c_+ c_-}{4N} \,  \frac{A_\perp} {A_z} \frac{A_\perp^3} {\omega_n^2}\\
  &\times \int_0^1 dx \int_0^1 dy \, \frac{x (\log x)^2 \, y (\log y)^2}{s+i (x-y)}
  \label{eq:sigma4cont}
\end{align}
in the continuum limit,
where $x=\exp\{-r_1^2\}$, $y=\exp\{-r_2^2\}$, and $r_i = \sqrt{x_i^2+y_i^2}/L$ ($i=1,2$).
Here, we have approximated $|A_k^\pm|^2 \simeq |A_{k,\mathrm{CB}}^\pm|^2 + |A_{k,\mathrm{LH}}^\pm|^2$ since the
overlap term vanishes under spatial averaging.
The appearance of polynomial prefactors $r^4$, represented by the $\log$ functions in the numerator of Eq. \eqref{eq:sigma4cont},
is a direct consequence of the excited-state envelope functions describing the distribution
of transverse coupling constants $A_k^\pm$ within the quantum dot.

The transverse-spin dynamics of the HH are described by the non-analytic structure of the right-hand side of Eq.
\eqref{eq:GME-Laplace} (see, e.g., Ref. \cite{coish:2010}).
Inserting $\Sigma^{(2)}$ and $\Sigma^{(4)}$ into Eq. \eqref{eq:GME-Laplace}, we find one 
pole at $s \simeq i \Delta \omega - \Gamma$, whose negative real part gives the HH decoherence rate 
$\Gamma = 1/T_2 \simeq - \mathrm{Im} \Sigma^{(4)}(i \omega_n + i \Delta \omega - 0^+)$ \cite{coish:2010},
where $0^+$ denotes a positive infinitesimal. Evaluating Eq. \eqref{eq:sigma4cont}, 
we find (see Appendix \ref{appendix2})
\begin{multline}
  \label{eq:Gamma}
  \frac{1}{T_2} = \frac{\pi c_+ c_-}{4N} \, \frac{A_\perp} {A_z} \frac{A_\perp^3} {\omega_n^2} \\
  \times \int_\epsilon^1 dx \, x [\log x]^2 (x+\epsilon) [\log(x+\epsilon)]^2,
\end{multline}
where $\epsilon = N | \Delta \omega  / A_{\mathrm{HH}} |$. The integral in Eq. \eqref{eq:Gamma} can now be evaluated
numerically for any value of $\epsilon$.

\begin{figure}[t]
  \centering
  \includegraphics[width=.98\columnwidth]{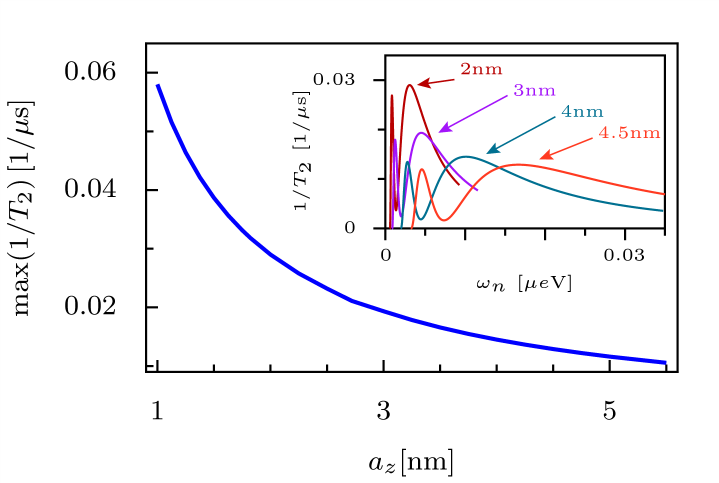}
  \caption{Maximum of the decoherence rate $1/T_2$ as a function of the QD height $a_z$. 
    For increasing $a_z$, the maximal value of $1/T_2$ decreases (see main plot) 
    and the position of the maximum is shifted (see inset).}
  \label{fig:Gammamax-of-az}
\end{figure}

The Zeeman energy of the HH is given by $\omega_n = g_h \mu_B B + p I A_{\mathrm{HH}}$, where $-1 \leq p \leq 1$ is the degree
of nuclear-spin polarization (along the positive $z$-direction).
In Fig. \ref{fig:Gamma-of-B}, we show the hole-spin decoherence rate $1/T_2$ as a function of $\omega_n$.
The non-monotonic behavior of $1/T_2$ for small $\omega_n$ appears when $\epsilon \propto 1/\omega_n$ 
approaches unity.

For electrons, a non-monotonic behavior of $1/T_2$ has been predicted as well \cite{coish:2010}, albeit 
with a different dependence on $\epsilon$ and around magnetic fields of several Tesla. 
In contrast, for holes, the non-monotonicity occurs at much lower fields ($B \simeq 0.1 \mathrm{mT}$ for the parameters 
used in Fig. \ref{fig:Gamma-of-B} assuming $p=0$), and the rate $1/T_2$ features an additional dip which is a footprint of the excited-state envelope functions
appearing in Eq. \eqref{eq:spinstates}. The huge difference in energy scales 
has very important consequences for the tunability of the hole-spin decoherence rate:
by increasing the externally applied magnetic field (or the degree of nuclear-spin polarization), 
it is possible to decrease $1/T_2$ over many orders of magnitude
within the experimentally accessible range of magnetic fields (see inset of Fig. \ref{fig:Gamma-of-B}).
This means that this system offers the possibility to entirely `turn off' hyperfine-associated spin decoherence.
As a consequence, hole-spin quantum dots may be operated in a regime where other interactions, such as
spin-orbit or direct nuclear dipole interactions, will be the dominant source of spin decoherence and will therefore
become experimentally observable. On the other hand,
for small $\omega_n$, the hybridization-induced transverse
interaction can be expected to be the dominant source of hole-spin decoherence.
We emphasize that Eq. \eqref{eq:Gamma} is still valid at $B=0$, as long as $\epsilon < 1$. For $\epsilon > 1$, 
$\Delta \omega$ exceeds the bandwidth of excitations $A_z/N$ in the nuclear bath, and the perturbation expansion breaks down
 \cite{coish:2010}.

The degree of band hybridization, and therefore the decoherence rate $1/T_2$, depends on the geometry of the QD,
i.e., on $L$ and $a_z$. 
For flat QDs the amount of LH admixture to the HH states \eqref{eq:spinstates} is decreased, leading to smaller non-Ising terms 
in the Hamiltonian \eqref{eq:holehamiltonian}. On the other hand, the envelope wavefunction of a flat dot encloses
less nuclear spins (for fixed $L$).
These two effects lead to an increase of the maximal decoherence rate for smaller $a_z$ 
(see Fig. \ref{fig:Gammamax-of-az}), 
and a shift of its position as a function of $\omega_n$ (see inset of Fig. \ref{fig:Gammamax-of-az}).

We acknowledge discussions with B. Braunecker, W.~A. Coish,  F. Pedrocchi, D. Stepanenko, and M. Trif, and funding
from the Swiss NSF, NCCR Nanoscience, and DARPA Quest.

\appendix

\section{Band hybridization \label{appendix1}}

Here we show how to derive the hybridized heavy-hole states given in Eq. (1) of the main text.
We start from the bulk version of the Kane Hamiltonian, which reads \cite{winkler}
\begin{equation}
  \label{supp:kane2}
  H_K = 
  \left( \begin{array}{cccc}
      H_{\mathrm{CB}} & V_1 & V_2 & V_3\\
      V_1^{\dagger} & H_{\mathrm{HH}} & V_4 & V_5\\
      V_2^{\dagger} & V_4^{\dagger} & H_{\mathrm{LH}} & V_6\\
      V_3^{\dagger} & V_5^{\dagger} & V_6^{\dagger} & H_{\mathrm{SO}}
    \end{array} \right),
\end{equation}
where the relevant blocks are given by
\begin{equation}
  \begin{array}{ll}
  H_{\mathrm{HH}} = \begin{pmatrix} B & 0\\ 0 & B\end{pmatrix},&
  V_1 = \frac{1}{\sqrt{2}} \begin{pmatrix} -E & 0\\ 0 & E^*\end{pmatrix},\\
  V_4 = \begin{pmatrix} 2 \sqrt{3} \, J & \sqrt{3} \, F\\ \sqrt{3} \, F^* & -2 \sqrt{3} \, J^* \end{pmatrix},&
  V_5 = \begin{pmatrix} -\sqrt{6} \, J & -\sqrt{6} \, F\\ \sqrt{6} \, F^* & -\sqrt{6} \, J^* \end{pmatrix},
  \end{array}
\end{equation}
with
\begin{align}
  \nonumber B &= -\epsilon [(\gamma_1 + \gamma_2) (k_x^2 + k_y^2) + (\gamma_1 - 2\gamma_2) k_z^2],\\
  \nonumber E &= P k_+,\\
  \nonumber F &= \epsilon [\gamma_2 (k_x^2 - k_y^2) -2i \gamma_3 k_x k_y ],\\
  J &= \epsilon \gamma_3 k_- k_z.
\end{align}
Here, $\epsilon = \hbar^2/2m_0$, $m_0$ is the free-electron mass, ${k_\pm = k_x \pm i k_y}$,
$\gamma_j$ denote the Luttinger parameters, and $P$ is the inter-band momentum.
$H_{\mathrm{CB}}$, $H_{\mathrm{LH}}$, and $H_{\mathrm{SO}}$ describe unperturbed electron states in the CB,
LH and SO bands, while $V_2$, $V_3$, and $V_6$ describe the CB-LH, CB-SO, and LH-SO
coupling, respectively, and thus do not contribute to the first order of the perturbation expansion
carried out in Eq. \eqref{holes:expansion} below.
We have neglected terms that are proportional to $C$, $B_{7v}$, and $B_{8v}^\pm$ in Winkler's notation \cite{winkler}
since, due to their smallness, they will not lead to considerable corrections for our purposes. 

We choose a parabolic confinement potential 
\begin{multline}
  V_{\mathrm{dot}} = V_z(z) + V_{xy}(x,y),\\  V_z(z) = \frac{m_\perp \omega_\perp^2}{2} z^2, \quad
  V_{xy}(x,y) = \frac{m_\| \omega_\|^2}{2} (x^2+y^2),
\end{multline}
where $\omega_\perp = \hbar/(m_\perp a_z^2)$ and $\omega_\| = \hbar/(m_\| L^2)$ with HH effective masses
$m_\perp = m_0 / (\gamma_1 - 2 \gamma_2)$ and $m_\| = m_0/(\gamma_1 + \gamma_2)$ along the growth direction
and in the plane of the dot, respectively, and $a_z$ and $L$ denote the corresponding confinement lengths.

In order to calculate the hybridized HH states perturbatively, we take the ground-state eigenket
$|\phi_{\mathrm{HH}}^{00} \rangle$ of
$H_{\mathrm{HH}} + V_{\mathrm{dot}}$ as the unperturbed envelope function, whose position representation is given by
\begin{equation}
  \langle \mathbf{r} | \phi^{00}_{\mathrm{HH}} \rangle = \phi^{0 \perp}_{\mathrm{HH}}(z)\phi^{0 \|}_{\mathrm{HH}}(x,y),
\end{equation}
with harmonic-oscillator eigenfunctions
\begin{align}
  \phi^{0 \perp}_{\mathrm{HH}}(z) &= \frac{1}{\sqrt{\sqrt{\pi} a_z}} \exp \left\{ -\frac{1}{2} \, \frac{z^2}{a_z^2}\right\},\\
  \phi^{0 \|}_{\mathrm{HH}}(x,y) &= \frac{1}{\sqrt{\pi} L} \exp \left\{ -\frac{1}{2} \, \frac{x^2+y^2}{L^2}\right\}.
\end{align}
Recall that the position representation of the envelope functions in band $\alpha$ is given by
$\langle \mathbf{r} | \phi_\alpha^{ij} \rangle = \phi_\alpha^{i \perp}(z) \phi_\alpha^{j \|}(x,y)$ ($i=0,1$, $j=0,\pm$), 
where $\phi_\alpha^{0 \|}(x,y)=\phi_\alpha^0(x) \phi_\alpha^0(y)$, $\phi_\alpha^{\pm \|}(x,y) = 
(\phi_\alpha^1(x) \phi_\alpha^0(y) \pm i \phi_\alpha^0(x) \phi_\alpha^1(y))/\sqrt{2}$,
and $\phi_\alpha^n(x)$ is the $n^\mathrm{th}$ harmonic-oscillator eigenfunction in band $\alpha$.
The hybridized HH states can now be evaluated by a perturbation expansion (in inverse energy splitting of
the bands) as follows:
defining a two-spinor $|\Psi_{\mathrm{HH}} \rangle=(|\phi^{00}_{\mathrm{HH}} \rangle, |\phi^{00}_{\mathrm{HH}} \rangle)^t$,
the hybridized HH spin states are the components of
\begin{align}
  \nonumber
   |\Psi_{\mathrm{hyb}} \rangle = \sum_{l,m,n} \biggl( &-\frac{1}{E_g} \, |\Psi^l_{\mathrm{CB}} \rangle \langle \Psi^l_{\mathrm{CB}}| \, V_1 \\
     &+ \frac{1}{\Delta_{\mathrm{LH}}} \, |\Psi^m_{\mathrm{LH}} \rangle \langle \Psi^m_{\mathrm{LH}}| \, V_4^\dagger \label{holes:expansion}\\
   &+ \frac{1}{\Delta_{\mathrm{LH}}+\Delta_{\mathrm{SO}}} \, |\Psi^n_{\mathrm{SO}} \rangle \langle \Psi^n_{\mathrm{SO}}| \,
   V_5^\dagger \biggr) |\Psi_{\mathrm{HH}} \rangle, \nonumber
\end{align}
where $l$, $m$, and $n$ label the electronic states in the CB, LH, and SO band, respectively,
$E_g$ is the fundamental band gap, $\Delta_{\mathrm{LH}}$ is the HH-LH splitting (see Eq. \eqref{delta-so}), and
$\Delta_{\mathrm{SO}}$ is the spin-orbit-induced LH-SO band splitting.

Due to the cylindrical
symmetry of the dot, only one matrix element per band yields a non-vanishing contribution to the sum.
For instance, $\langle \phi^{00}_{\mathrm{CB}} | k_\pm | \phi^{00}_{\mathrm{HH}} \rangle = 0$ due to the vanishing
angular integral, but $\langle \phi^{0 \pm}_{\mathrm{CB}} | k_\pm | \phi^{00}_{\mathrm{HH}} \rangle \neq 0$,
where $\langle \mathbf{r} | \phi^{0 \pm}_{\mathrm{CB}} \rangle = \phi^{0 \perp}_{\mathrm{CB}}(z) \phi^{\pm \|}_{\mathrm{CB}}(x,y)$
with
\begin{equation}
  \phi^{\pm \|}_{\mathrm{CB}}(x,y) = \frac{x \pm i y}{L'} \phi^{0 \|}_{\mathrm{CB}}(x,y).
  \label{eq:excitedstate}
\end{equation}
$L' = L \alpha/\sqrt{\gamma_1+\gamma_2}$ is the renormalized lateral confinement length of a
CB electron (due to the difference in effective masses between particles in the CB and HH band), 
where $\alpha = \sqrt{m_0/m^*}$ with the free-electron mass $m_0$ and the effective CB electron mass $m^*$.
The excited-state envelope function \eqref{eq:excitedstate} features a polynomial prefactor which gives rise to the
appearance of logarithmic terms in the self-energy \eqref{appendix:self-energy-log} and,
eventually, to a minimum of the decoherence rate $1/T_2$ as a function of $\omega_n$
(see Fig. 1 of the main text).
The matrix element
\begin{equation}
  \beta_{\mathrm{CB}} = \langle \phi^{0 \pm}_{\mathrm{HH}} | \phi^{0 \pm}_{\mathrm{CB}} \rangle =
  \frac{4 \alpha (\gamma_1+\gamma_2)}{(\alpha+\gamma_1+\gamma_2)^2} \frac{\sqrt{2} (\alpha(\gamma_1-2\gamma_2))^{1/4}}
  {\sqrt{\alpha+(\gamma_1-2\gamma_2)}}
\end{equation}
differs from unity because of the different effective CB and HH masses which lead to different
spring constants of the harmonic-oscillator potential in the CB and HH bands.
Similarly, for the LH contribution, the only non-vanishing term in the sum in Eq. \eqref{holes:expansion} is given by
$\langle \phi^{1 \pm}_{\mathrm{LH}} | k_\pm k_z | \phi^{00}_{\mathrm{HH}} \rangle$, and the overlap integral reads
\begin{equation}
  \beta_{\mathrm{LH}} = \langle \phi^{1 \pm}_{\mathrm{HH}} | \phi^{1 \pm}_{\mathrm{LH}} \rangle =
  \left[ 1 - \left( \frac{\gamma_2}{\gamma_1} \right)^2 \right] \frac{(\gamma_1^2 - 4\gamma_2^2)^{3/4}}{\gamma_1^{3/2}}.
\end{equation}

The HH-LH splitting $\Delta_{\mathrm{LH}}$ is given by the difference in ground-state energies in the HH and
LH bands: 
$\Delta_{\mathrm{LH}} = E_{\mathrm{HH}} - E_{\mathrm{LH}}$,
where
\begin{align}
  E_{\mathrm{HH}} &= -\frac{\hbar^2}{2} \, \left( 2 \frac{\gamma_1+\gamma_2}{m_0 L^2} 
  + \frac{\gamma_1 - 2 \gamma_2}{m_0 a_z^2} \right), \\
  E_{\mathrm{LH}} &= -\frac{\hbar^2}{2} \, \left( 2 \frac{\gamma_1-\gamma_2}{m_0 L^2} 
  + \frac{\gamma_1 + 2 \gamma_2}{m_0 a_z^2} \right).
\end{align}
This leads to a HH-LH splitting of
\begin{equation}
  \Delta_{\mathrm{LH}} = \frac{2 \hbar^2 \gamma_2}{m_0} \,  \frac{L^2-a_z^2}{L^2 a_z^2}.
  \label{delta-so}
\end{equation}

Carrying out the perturbation expansion following Eq. \eqref{holes:expansion}, inserting
the quantities given above, and neglecting the small SO band contributions induced by $V_5$,
we arrive at the hybridized states given in Eq. (1) of the main text.
Note that the Kane Hamiltonian \eqref{supp:kane2} is written in the basis of
CB, HH, LH, and SO Bloch states, so that these Bloch states only appear implicitly in Eq. 
\eqref{holes:expansion}, but explicitly in the final form given in Eq. (1) in the main text.
In order to produce explicit numbers, the Bloch amplitudes have been approximated by a linear
combination of $n=4$ hydrogen-like atomic orbitals, such as in Ref. \cite{fischer:2008}.

We have also neglected HH-LH coupling terms induced by the $F$ matrix element in $V_4$ since such terms
are parametrically suppressed by a factor of $a_z/L \ll 1$ with respect to the couplings induced by the $J$
matrix element.
Note that the LH (pseudospin) Bloch functions near the $\Gamma$-point,
\begin{equation}
|u_{\mathrm{LH} \pm}\rangle |\pm_{\mathrm{LH}}\rangle \simeq ({\sqrt{2} |p_z\rangle |\uparrow, \downarrow \rangle} 
\mp {|p_\pm\rangle |\downarrow, \uparrow \rangle})/\sqrt{3},
\end{equation}
already include (real) spin-up and spin-down states, such that the $J$ matrix element, although diagonal
in pseudospin space, induces off-diagonal couplings in the effective Hamiltonian (Eq. (5) of the
main text). Therefore, the $F$ matrix element does not introduce qualitatively new contributions to 
the effective Hamiltonian.

\begin{figure}[t]
  \centering
  \includegraphics[width=.9 \columnwidth]{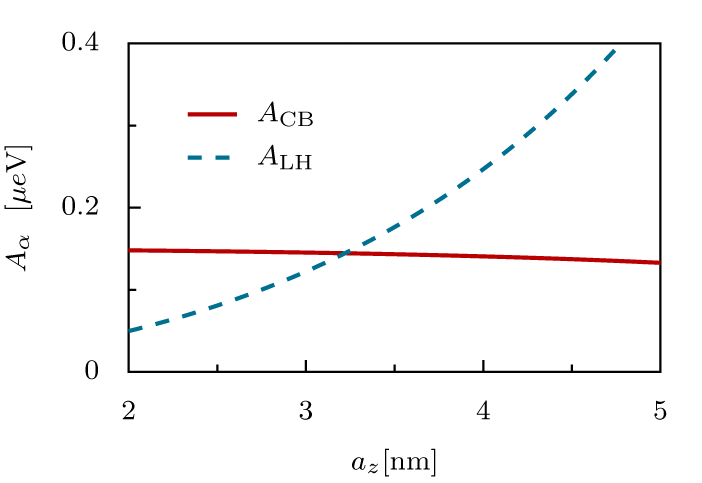}
  \caption{Conduction-band (CB) and light-hole (LH) contributions to the non-Ising hyperfine coupling $A_\perp$ as a function of the perpendicular dot size $a_z$. For this plot, we have chosen a GaAs dot of lateral size $L=10 \mathrm{nm}$.}
  \label{fig:coupling}
\end{figure}

The degree of band hybridization is described by the dimensionless prefactors
\begin{equation}
  \lambda_{\mathrm{CB}} = \frac{i \beta_{\mathrm{CB}} P}{\sqrt{2} L E_g}, \quad
  \lambda_{\mathrm{LH}} = \frac{\sqrt{3} \beta_{\mathrm{LH}} \gamma_3 a_z L}{2 \sqrt{2} \gamma_2 (L^2-a_z^2)}.
\end{equation}
We see that these prefactors, and therefore also the hyperfine coupling strengths 
$A_{\mathrm{CB}} \propto |\lambda_{\mathrm{CB}}|^2$ and $A_{\mathrm{LH}} \propto |\lambda_{\mathrm{LH}}|^2$, 
have a strong dependence on the QD geometry (i.e., on $L$
and $a_z$, therefore the non-Ising terms in Eq. (5) of the main text can, in principle, be `tuned' to be 
more CB or LH-like (see Fig. \ref{fig:coupling}).
It should be emphasized, however, that CB and LH admixture lead to qualitatively 
similar contributions to the effective Hamiltonian, Eq. (5) in the main text.
For a quantum dot with $L=10\mathrm{nm}$ and $a_z=2 \mathrm{nm}$, we estimate
$|\lambda_{\mathrm{CB}}| \simeq 0.04$, $|\lambda_{\mathrm{LH}}| \simeq 0.11$ for GaAs.

\section{Continuum limit \label{appendix2}}

Here we show in detail how to calculate the decoherence rate $\Gamma = 1/T_2$ from the fourth-order self-energy in the
continuum limit. We start from Eq. (9) of the main text,
\begin{equation}
  \Sigma^{(4)}(s+i \omega_n) \simeq -i \frac{c_+ c_-}{4 \omega_n^2} \sum_{k_1,k_2}
  \frac{|A_{k_1}^\pm|^2 |A_{k_2}^\pm|^2}{s+i (A_{k_1}^z - A_{k_2}^z)}.
  \label{appendix:sigma4discrete}
\end{equation}
We now replace sums by integrals according to $v_0 \sum_k \rightarrow \int d^3 r$, where $v_0$ is the volume
occupied by one nucleus and the integration is carried out over all space.
Averaging over the $z$-dependence of the coupling constants, $A(x,y) = \int_{-\infty}^{\infty} dz \, A(x,y,z)$
we write $\Sigma^{(4)}$ in terms of a four-dimensional spatial integral:
\begin{multline}
  \Sigma^{(4)}(s+i \omega_n) \simeq -i \frac{c_+ c_-}{4 \pi^2 N} \frac{A_\perp^4}{\omega_n^2 A_z}
  \int_0^\infty dr_1 r_1 \int_0^\infty dr_2 r_2\\ \times \int_0^{2 \pi} d \theta_1 \int_0^{2 \pi} d \theta_2 \,
  \frac{r_1^4 r_2^4 e^{-2r_1^2} e^{-2r_2^2}}{\mathsf{s}+i (e^{-r_1^2} - e^{-2r_2^2})},
\end{multline}
where we have introduced spherical coordinates $x_i = r_i \cos \theta_i$, $y_i = r_i \sin \theta_i$,
and where $\mathsf{s} = s N/A_z$. We have also rewritten $N v_0 = \pi a_z L^2$ (the volume of the QD).
The angular integrals simply contribute a prefactor of $4 \pi^2$.
The radial integrals can be solved by introducing new variables $x = e^{-r_1^2}$, $y = e^{-r_2^2}$, such that
\begin{multline}
  \Sigma^{(4)}(s+i \omega_n) \simeq -i \frac{c_+ c_-}{4 N} \frac{A_\perp^4}{\omega_n^2 A_z}\\ \times
  \int_0^1 dx \int_0^1 dy \, \frac{x (\log x)^2 \, y (\log y)^2}{\mathsf{s}+i (x-y)}.
  \label{appendix:self-energy-log}
\end{multline}

The right-hand side of the equation of motion in Laplace space (Eq. (7) of the main text),
\begin{equation}
  S^+(s+i \omega_n) = \frac{\langle S^+ \rangle_0}{s +i \Sigma(s+i \omega_n)}
\end{equation}
features a pole at $s = i\Delta \omega - \Gamma$ \cite{coish:2008, coish:2010},
where $\Gamma = -\mathrm{Im} \Sigma^{(4)} (i \omega_n + i \Delta \omega - 0^+)$
and where $0^+$ denotes a positive infinitesimal.
Evaluating $\Sigma^{(4)}$ at $s=i \omega_n + i \Delta \omega - 0^+$ and using
\begin{equation}
  \lim_{\eta \rightarrow 0} \frac{1}{\xi \pm i \eta} = \mathcal{P} \frac{1}{\xi} \mp i \pi \delta(\xi),
\end{equation}
where $\mathcal{P}$ denotes that the principle value should be taken in any integration involving the above expression,
we arrive at an integral of the form
\begin{multline}
  I = \int_0^1 dx \int_0^1 dy \, f(x,y)\\ \times \left( \mathcal{P} \frac{1}{x-y+\Delta \omega} - i \pi \delta (x-y+\Delta \omega) \right).
\end{multline}
Taking the imaginary part according to
$\Gamma = -\mathrm{Im} \Sigma^{(4)} (i \omega_n + i \Delta \omega - 0^+)$ leads directly to Eq. (12)
of the main text.

\end{document}